\documentclass[a4paper,11pt]{article}
\usepackage{amssymb,amsmath}
\usepackage{enumitem}
\usepackage{graphicx,graphics}
\usepackage{times}
\usepackage[english]{babel}
\usepackage[colorlinks=true, urlcolor=blue, linkcolor=black, citecolor=black]{hyperref}

\newcommand{\BE}{\begin{equation}}
\newcommand{\EE}{\end{equation}}
\newcommand{\A}{\boldsymbol{A}}
\newcommand{\B}{\boldsymbol{B}}
\renewcommand{\b}{\boldsymbol{b}}
\newcommand{\m}{\boldsymbol{m}}
\renewcommand{\r}{\boldsymbol{r}}
\newcommand{\y}{\boldsymbol{y}}
\newcommand{\z}{\boldsymbol{z}}
\newcommand{\G}{\mathcal{G}}
\renewcommand{\L}{\mathcal{L}}
\renewcommand{\P}{\mathcal{P}}
\newcommand{\Q}{\mathcal{Q}}
\newcommand{\R}{\mathcal{R}}
\newcommand{\ds}{\displaystyle}
\newcommand{\ovl}{\overline}
\newcommand{\arc}[1]{\lower-4pt\hbox{${\displaystyle\frown\atop{\displaystyle #1}}$}}
\newcommand{\seg}[1]{\lower-4pt\hbox{${\displaystyle\atop{\displaystyle #1}}$}}

\usepackage{psfig}

\title{Numerics on the trajectory of nodule displacements by external compressions of the breast}

\begin{document}

\maketitle

\pagestyle{myheadings}

\centerline{Marcelo Zanchetta do Nascimento}

\centerline{\small\em FACOM, Federal University of Uberl\^andia, Uberl\^andia-MG, Brazil}

\centerline{Val\'erio Ramos Batista}

\centerline{\small\em CMCC, Federal University of ABC, Santo Andr\'e-SP, Brazil}

\hrulefill

\begin{abstract}
We present a fast and reliable algorithm that gives precise location of breast tumours for a partial mastectomy. Our algorithm is fully implemented in the Surface Evolver, which is a general-purpose simulator of physical experiments. By starting from the X-rays images that show a tumour one takes its 2D coordinates in each view. These views are called CC (Craniocaudal) and MLO (Mediolateral Oblique). Together with some measurements of the patient's breast, that coordinates are given as input to our simulator. From this point on the simulator reproduces all main steps of taking mammography with a virtual transparent breast that matches the patient's. The virtual mammography procedure is graphically displayed on the computer screen, so that users can track the virtual tumour inside the breast. As output we have the coordinates of the tumour position when the woman lies on the operating table for the surgery. With these coordinates the surgeon can make a small incision into the breast and reach the tumour for its removal. After a simple plastic correction the whole structure of the breast will be preserved.
 
\vspace{0.5cm}
\end{abstract}
\ \\

\hfill{\footnotesize{The 2nd author dedicates this work to his wife Clarice.}}



\section{Introduction}
\label{intro}
It was in 1981 that the renowned oncologist Umberto Veronesi introduced a technique of mastectomy called {\it quadrantectomy} (see \cite{Vero}). This technique consists of estimating the location of a breast tumour within one quarter of the structure, and then proceed with a partial mastectomy to remove only the quadrant that contains the tumour. Afterwards, the residual parenchyma of the breast is submitted to radiotherapy in order to hinder a {\it locally recurrent malignancy}. This kind of malignancy happens when the tumour reappears next to where it had been removed (for instance, in the surgery scars).

His technique was innovative because it challenged the general consensus that breast cancer could only be eradicated by a full mastectomy. Such a consensus was based on the high chance of occurrence of locally recurrent malignancy. The other kind of malignancy is called {\it metastasis}, when the cancer spreads to other parts of the body even not directly connected to its original location. It happens because some cancer cells may go into lymph or blood, and so they travel along the body until fixing in another tissue.

Umberto Veronesi's technique proved to be successful, though recommendable only in the case of relatively small tumours. He is the pioneer of partial mastectomy, also called {\it lumpectomy}, or even Breast-Conserving Surgery (BCS). Along the decades BCS has been developed towards reducing the size of breast portion for removal to a minimum. Of course, this minimum depends on finding the exact position of the tumour inside the breast. Many softwares have been developed for this purpose. Among others we cite \cite{BS,B,H,Ku,P} and also two nice surveys about them \cite{N,T}. Unfortunately, they still present relevant limitations already discussed in \cite{Zanka_icm2_2014,Zanka_icm2_2015}. To the best of our knowledge, until now, not one of them has been officially approved by a Medical Council as a reliable nodule locator permitting it to become part of surgical preparations.

Anyway, no matter how accurately we can locate a nodule the surgeon will always remove some tissue that surrounds it. This prevents from {\it infiltration}, which is an abnormal accumulation of cells around the cancer. By the way, the reader must have noticed that we use the terms cancer, tumour and nodule as if they were interchangeable. Without going into technical details, in this paper a nodule is a just a lump, a tumour is a nodule to be surgically removed, and a cancer is a malignant tumour. The so-called benign tumours can sometimes just remain under observation, but removal is frequently prescribed even in their case.

This work concludes a long-term research devoted to the first steps of a mammography simulator that we have been developing since 2012. In \cite{Zanka_icm2_2014,Zanka_icm2_2015} the reader will find some discussion about the state of the art regarding {\it virtual mammographies}. There we cite softwares and works on the subject, evaluate their weak and strong points, and give detailed information about our method. Extended versions of \cite{Zanka_icm2_2014} and \cite{Zanka_icm2_2015} are \cite{Zanka_rpmi} and \cite{Zanka_ijser}, respectively. 

We have followed an approach that differs by three main characteristics from what is already found in the literature: 1) our method utilizes the Surface Evolver \cite{Brakke}; 2) it includes the study of nodule displacements in transparent breast phantoms; 3) it achieves a fast, reliable and relatively simple algorithm thanks to the application of many geometrical and physical properties.  

Herewith we present our {\it reverse procedure} to locate nodules precisely: by starting from CC and MLO views that show a tumour we mark it on a layer in the virtual breast. The layer is then tested in order to confirm that the virtual tumour will reach the same positions showed in the X-ray images. One can apply small variations in both the layer and the marked tumour, which will be tracked by our software. In case of matching it will locate the tumour at the {\it surgery position}, that we abbreviate to SRG.

The location is given in polar coordinates $(r,p,d)$ centred at the nipple (see Figures 1 and 2). The letters {\it r, p} and $d$ stand for geodesic radius, phase and depth, respectively. For instance, with an eye pencil one can draw a coordinate system on the breast.

\centerline{
\includegraphics[scale=0.32]{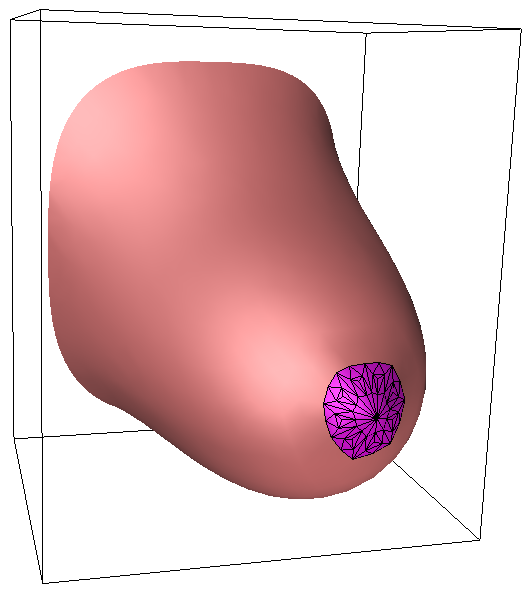}\hfill
\includegraphics[scale=0.32]{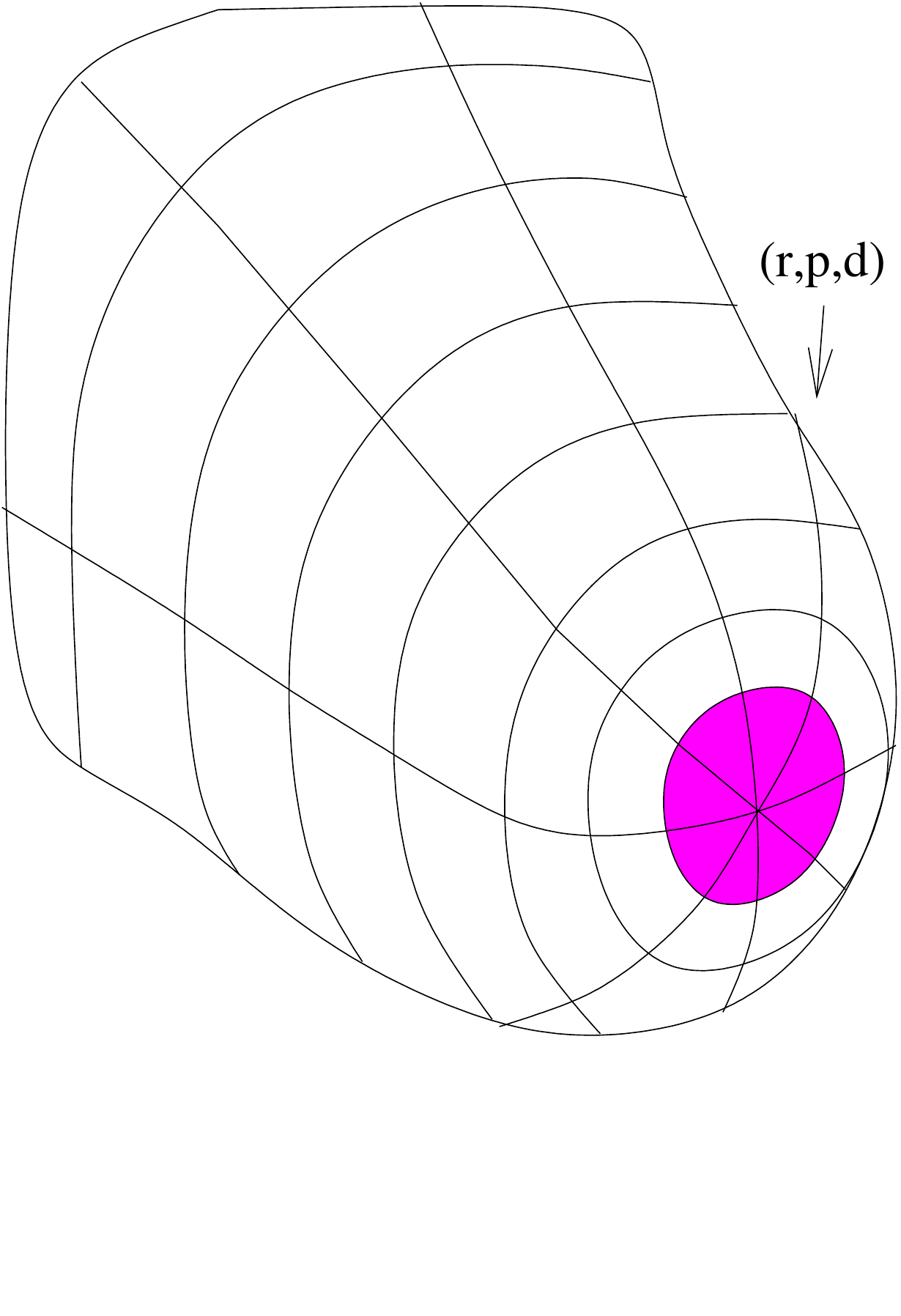}}
\centerline{Figure 1: Patient's breast.\hfill Figure 2: Drawing polar coordinates on the breast.}
\ \smallskip

Although the X-ray images are 2D-projections they can determine the {\it spatial} coordinates of the nodule as we shall explain in the next sections. Once the coordinates are known with a good precision at SRG the surgeon will then reach the nodule by starting at $(r,p)$ and cutting with the scalpel until depth $d$.

In case the tumour is found this will validate our modelling and also prove that the simplifications we have been adopting are acceptable. Otherwise the surgery will follow standard procedures and our modelling will then need further improvements like addition of more complexity, experiments with a wider variety of breast phantoms and software re-structuring.

\section{Preliminaries}
\label{prelim}

Of course, the polar coordinates depicted in Figure 2 cannot be registered in the X-ray images. But if a tumour is detected they will be helpful to determine the value of {\tt lf}, which is an essential parameter in our virtual mammography. The virtual mammography has 6 main steps named SRG ({\it surgery}), STU ({\it stand-up}), LAT ({\it lay-on-table}), CRC ({\it cranio-caudal}), LET ({\it lean-on-table}) and MLO ({\it medio-lateral-oblique}). Since we can use a {\it table} to take measurements of the patient's breast we prefer that term instead of {\it plate}. This one will only appear when the mammographer is really necessary. 

As commented in \cite{Zanka_icm2_2015}, since Evolver works with surface layers our approach is to mark a virtual nodule on a layer inside our model and track its trajectory. In Evolver we represent it by a black triangle. See Figures 3 and 4.
\\

\centerline{
\includegraphics[scale=0.44]{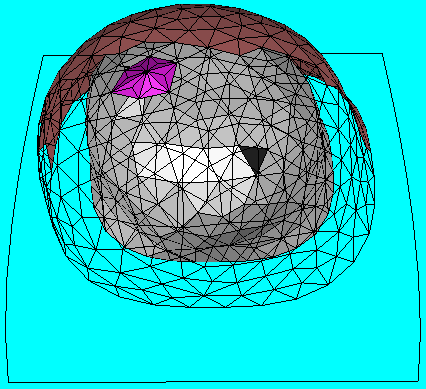}\hfill
\includegraphics[scale=0.35]{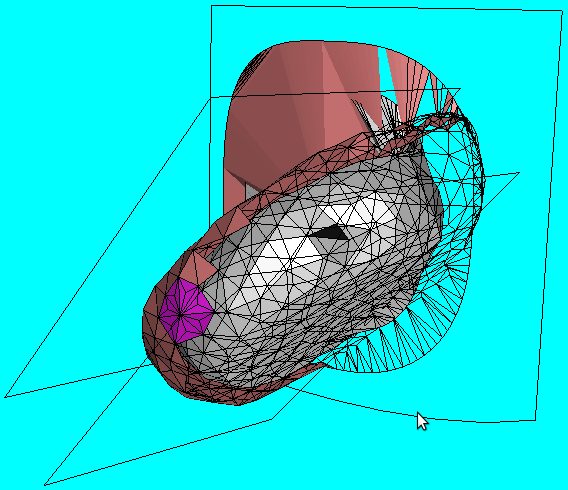}}

\centerline{Figure 3: Marking a nodule at SRG.\hfill Figure 4: Its position at MLO.}
\ \smallskip

We consider the layers as a reasonable approach because a tumour cannot move around independently of the {\it component} it is attached to: fat, gland, lactiferous duct, etc. Although inexact it is worthwhile to work with the layer-approach before increasing complexity in our model. 

The parameter name {\tt lf} is a mnemonic for ``layer factor'' and it indicates how close the nodule is to the breast skin. We have $0\le$ {\tt lf} $<1=$ 100\% because a nodule can never be on the skin. But an {\tt lf} close to 100\% means that it is almost there. An {\tt lf} almost zero means that the nodule appears close to the breast base in {\it both} CC and MLO views. Soon we shall give details about reaching a good estimate of {\tt lf}.

Once a tumour is found in the mammographies, and the gynaecologist concludes that a mastectomy is really necessary, the first thing they will need in order to locate the tumour by our method is to find {\tt lf}. Now we explain how to do it.

Figure 5 shows the breast at SRG. In \cite{Zanka_icm2_2014} we describe the correct way to reproduce this position: the woman lies down as if on a real operating table and holds a support fitted under her armpit. Then strategic measurements of her breast must be taken with a tape measure which, however, cannot be tightened around the parts. Such measurements are detailed in \cite{Zanka_icm2_2014,Zanka_icm2_2015}.

Afterwards one must draw the polar coordinates. The first thing to be drawn is the contour $bbcrc$ explained in \cite{Zanka_icm2_2015}. Next, ones draws the direction $Oy$ according to the way we position the $Oxyz$-coordinate system upon the breast (see \cite{Zanka_icm2_2014,Zanka_icm2_2015}). We remark that $Oy$ should approach a midline of the breast contoured by $bbcrc$, and should also pass close to the nipple. Since the breast is not perfectly symmetric $Oy$ will just balance both conditions. Anyway, it must be orthogonal to $bbcrc$. 

Now $Ox$ will pass through the nipple orthogonally to both $Oy$ and $bbcrc$. Together, $Ox$ and $Oy$ are called ``main directions''. After that we mark the diagonals $y=\pm\,x$ and the bisector between each diagonal and main direction. In this way we shall have divided the $360^\circ$-turn of $bbcrc$ in 16 sectors of $22.5^\circ$ each. Notice that all 16 rays must cross $bbcrc$ orthogonally.

Next we draw a contour that is parallel to $bbcrc$ and crosses each of those rays orthogonally at the respective halfway. Finally one draws another two parallel contours that cross the halfway of each resultant halving from the previous step. Eventually each ray will be equally divided in four parts (see Figure 6). Figure 6 also shows how to apply $(r,p,d)$ before using the scalpel. Once the point is found the scalpel will cut till depth $d$.

From this point on we present a qualitative analysis for a good estimate of {\tt lf}. Herewith we choose the left breast for explanations, which will be analogous for the right breast. A quantitative analysis will be discussed in the next section.

Figure 7 illustrates how the polar coordinates deform from SRG to either CRC or MLO. Of course, in practice the asymmetry of CRC can be closer to SRG than what is shown in the scheme. The $z$\,-level lines of MLO can also be less curved too. In fact, all variations will depend on the density and dimensions of the breast. For a preciser schematization one should observe what happens in each patient's case.
\\

\centerline{
\includegraphics[scale=0.50]{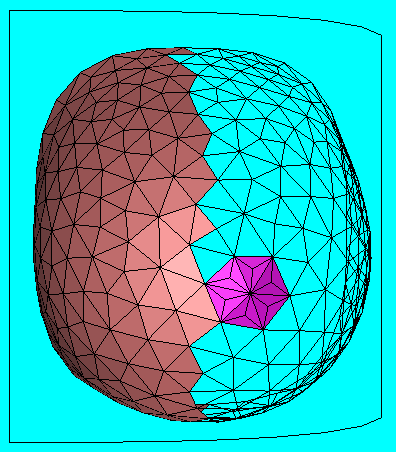}\hfill
\includegraphics[scale=0.50]{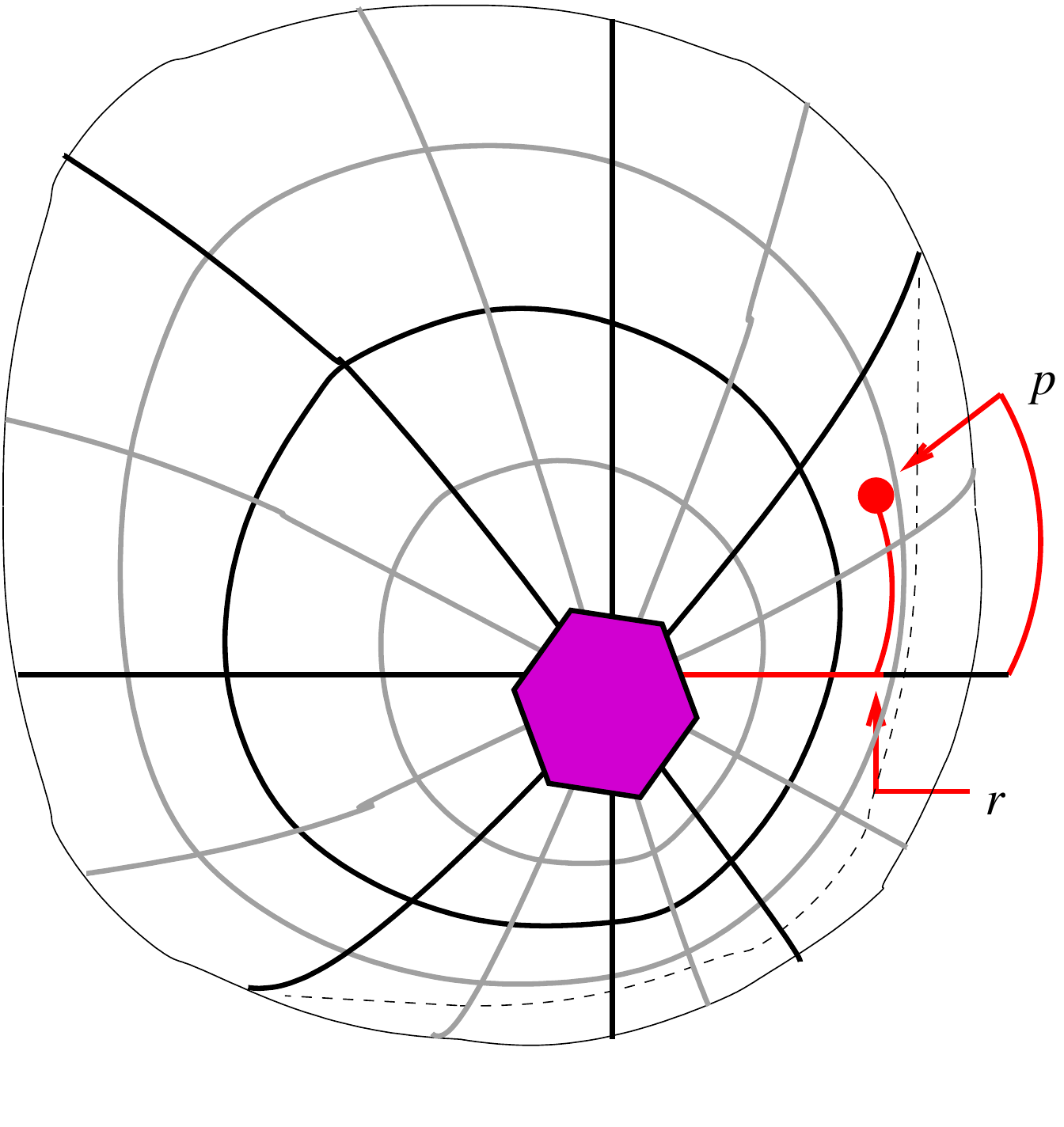}}

\centerline{Figure 5: Breast at SRG.\hfill Figure 6: Drawing and finding the point for the scalpel.}
\ \smallskip

Notice that ``CRC'' is our way to abbreviate the Craniocaudal {\it Compression}, not the X-ray CC\,-view. Namely, CRC is the three-dimensional state observed during the real (or virtual) mammography procedure. For ``MLO'' we just leave this difference to the context.
\\

\centerline{
\includegraphics[scale=0.55]{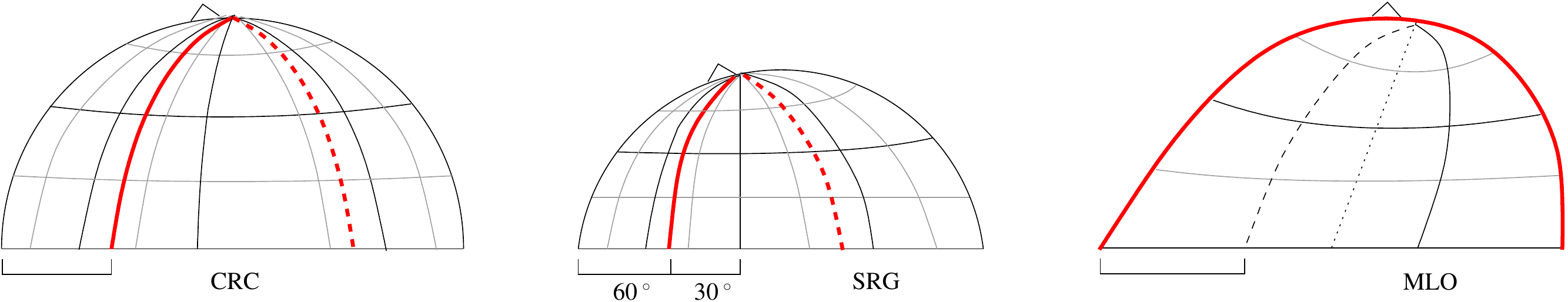}}

\centerline{Figure 7: Polar coordinates at strategic steps (patient's viewpoint orthogonal to the plate).}
\ \smallskip

In Figure 7 one sees a red line which will be parallel to the plate at MLO. Although the plate is tilted by 45$^\circ$ degrees the breast falls sideways when the woman is standing. This deforms the polar coordinates depicted in Figure 6 and so the ray $y=x$ is {\it not} parallel to the plate at MLO. Instead we have another ray that is approximately $y=\sqrt{3}\,x$, indicated by the red colour in Figure 7.

Figure 8 left is a scheme of an X-ray CC\,-view of the breast, and some coloured dots exemplify 21 positions where a nodule can be detected. The red and violet bullets show the case of nodules that are {\it almost} on the skin. 
\\

\centerline{
\includegraphics[scale=0.70]{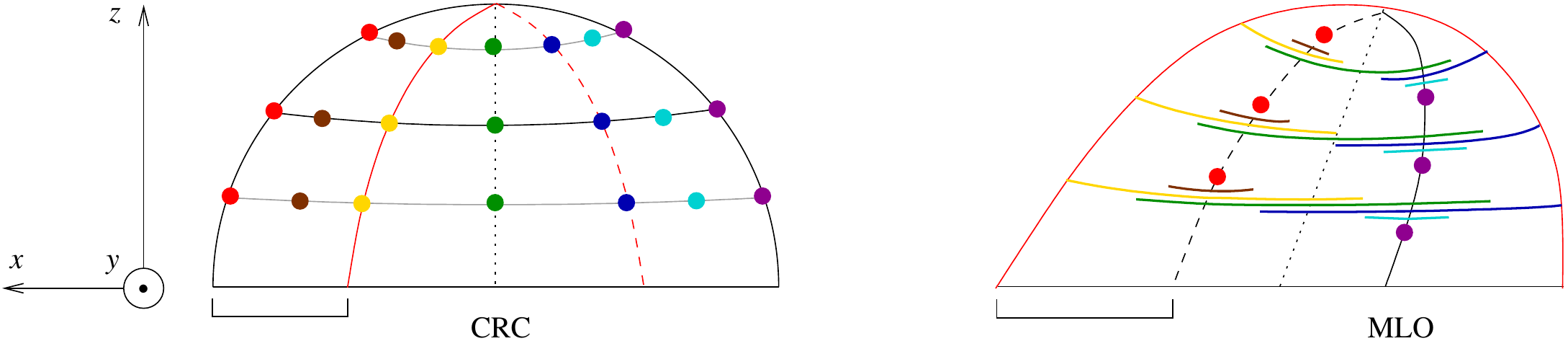}}

\bigskip

\centerline{Figure 8: Estimating nodule depth by means of CC and MLO views.}

\bigskip

What the CC\,-view cannot show is the nodule depth in the $y\,$-direction. Figure 8 right shows the MLO\,-view of the possible depths for each colour and $z\,$-level. Now let us take, for instance, the upper green line $\G$ of Figure 8 right. Since the CRC\,-scheme is (almost) symmetric by 180$^\circ$ rotations around the dotted line parallel to $Oz$, then $\G$ will be crossed in the middle by the corresponding dotted line at MLO. Moreover, by naming {\it equator} the arc at CRC that is contained in $Oxz$, then $\G$ will cross the equator twice at MLO. 

However, the extremes of $\G$ in Figure 8 are just figurative. In order to estimate their exact position one can make use of our simulator as we shall demonstrate in Sections~\ref{lnSRG} and \ref{appl}. Once the extremes are known the nodule that appears in the MLO view will lie at some point $\P\in\G$. For the moment assume that $\G$ corresponds to a straight line $G$ at SRG. Hence, if $\L$ and $\R$ are the left and right extremes of $\G$ in Figure 8 right, the corresponding extremes $L$ and $R$ of $G$ are viewed at the front and back of SRG, similarly to Figure 8 left. 

Now $\G=\arc{\L\R}$ and $G=\ovl{LR}$. Assume further that $\P$ corresponds to the $P\in G$ that verifies 
\BE
   \frac{\big|\seg{\ovl{LP}}\big|}
        {\big|\seg{\ovl{LR}}\big|}=
   \frac{|\arc{\L\P}|}{|\arc{\L\R}|}=:\rho,\label{rho}
\EE
where the symbol $=:$ means ``denoted''. Later we shall see that these assumptions have good evidence of being true. 

\section{Computing the Layer Factor}
\label{lfact}

By carrying on with our example from the previous section we recall that $\G$ is viewed as a single green dot in Figure 8 left. However, for the sake of generality we shall consider that the green dot is {\it not} exactly on $Oz$. It now lies at the same $z$-level but somewhere between the original position and the yellow dot. 

That $z$-level we denote by $\z$, which corresponds to $h=c\,\z\in Oz$ at SRG. The re-scaling constant $c$ depends only on the density of the breast.

As explained in \cite{Zanka_icm2_2014,Zanka_icm2_2015} we use the approach of taking the breast at SRG as an upper half-ellipsoid, but only to compute initial values. Our simulator will then give a more reliable shape by a semi-supervised method explained in that previous works.

Even as a half-ellipsoid its radii $x_r$, $y_r$ and $z_r$ will differ at most 11\% from each other. For instance, in \cite{Zanka_icm2_2014,Zanka_icm2_2015} we show an example where $y_r=z_r=7$ and $x_r=6.25$. Therefore, in order to find the layer factor {\tt lf} we shall consider the upper hemisphere in Figure 9.

There we have $|\ovl{OH}|=h$ and $|\ovl{OL}|=|\ovl{OR}|$ is the average $a=(x_r+y_r+z_r)/3$. Hence
\BE
   |\ovl{LR}|=2\cos\theta\sqrt{a^2-h^2},
\EE
where $\theta$ is indicated in Figure 9. Now the cosine law gives
\BE
   |\ovl{OP}|^2=
   |\ovl{LP}|^2+a^2-2a|\ovl{LP}|\cos\theta\sqrt{1-(h/a)^2}.
\EE
Finally, notice that
\BE
   \frac{|\ovl{LP}|}{a}\stackrel{(1)}{=}
   \rho\cdot\frac{|\ovl{LR}|}{a}\stackrel{(2)}{=}
   2\rho\cos\theta\sqrt{1-(h/a)^2}.
\EE

Our layer factor is then {\tt lf}\,$=|\ovl{OP}|/a$. From (4) we rewrite (3) as 
\BE
   {\tt lf}^2=1-4\rho(1-\rho)
   \biggl[1-\biggl(\frac{h}{a}\biggl)^2\biggl]\cos^2\theta.\label{lf}
\EE

\centerline{
\includegraphics[scale=0.45]{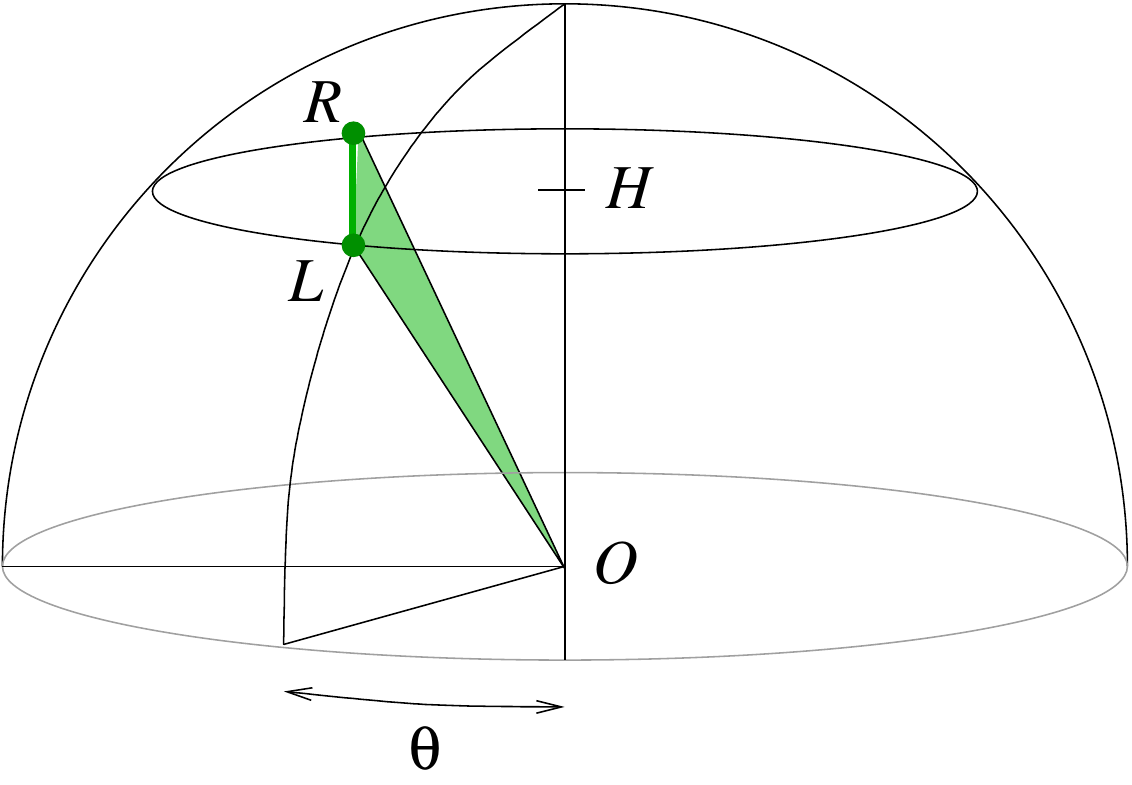}}

\centerline{Figure 9: Symmetrized breast at SRG.}

\ \smallskip

Of course, $\theta$ cannot be taken directly from the mammographies. Instead, one should observe how the polar coordinates deform at CRC. By looking at Figure 8 left there will be a line similar to the red one that passes through the image of the nodule in the CC\,-view. Since the red line corresponds to circa $30^\circ$ with respect to $Oy$, then one should deduce the approximate value of $\theta$. For instance, it could be quite close to the grey line at CRC in Figure 7, hence $\theta\simeq 22.5^\circ$.

However, deducing $\theta$ this way is not necessary for our simulations. The virtual mammographer will give $(r,p,d)$, where $p\simeq 90^\circ-\theta$ whenever ${\tt lf}\simeq 1$. Hence the surgeon can compare our simulator output with the deduced $\theta$ just to check consistency.

\section{Experiments with the Phantom}
\label{exphant}

As explained in \cite{Zanka_icm2_2015} we have performed experiments with a transparent breast phantom containing artificial nodules. For convenience of the reader we reproduce our results of that paper once again in Figures 10 and 11.

Now consider the $OXZ$\,-view of Figure 10. By labelling its nodules alphabetically from left to right and top to bottom one can read off the coordinates after and before CRC. They are all listed in Table 1.

We are going to see that the nodules displace according to a {\it linear function} if one takes $\y=Y+2.25$. Notice that we make use of different coordinate systems: $Oxyz$ for the simulator, $OXYZ$ for the phantom, and some others in bold style for tranforms. The choice of $Oxyz$ was made to be compatible with Evolver, whereas $OXYZ$ simplifies reading the grid paper.

Before the compression the eleven $(X,\y,Z)$ coordinates can be listed in a matrix $\B_{11\times 3}$. After compression the corresponding matrix is $\A_{11\times 3}$. We are looking for a matrix of coefficients $C_{3\times 3}$ that minimizes $E=\A-\B C$. More precisely, the sum of the squares of the $E$-entries must attain the least possible value. This is equivalent to 
\BE
   C=(B^tB)^{-1}B^tA=
   \left[
   \begin{array}{rrr}
   1.20 & -0.07 & 0.02\\
   0.02 &  1.31 & 0.12\\
  -0.05 &  0.00 & 1.15
   \end{array}
   \right].
\EE 

\centerline{
\includegraphics[scale=0.20]{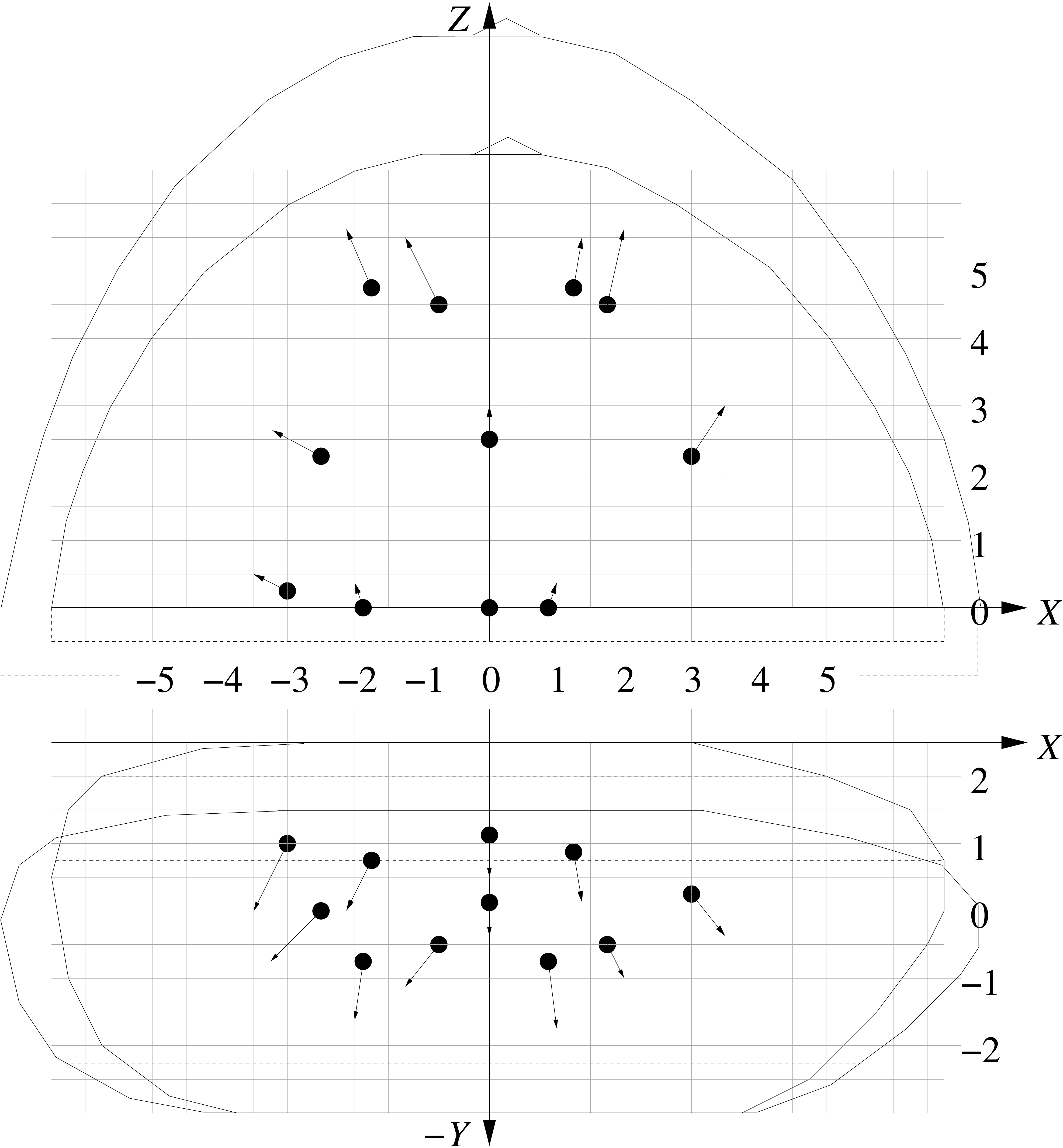}\hfill
\includegraphics[scale=0.20]{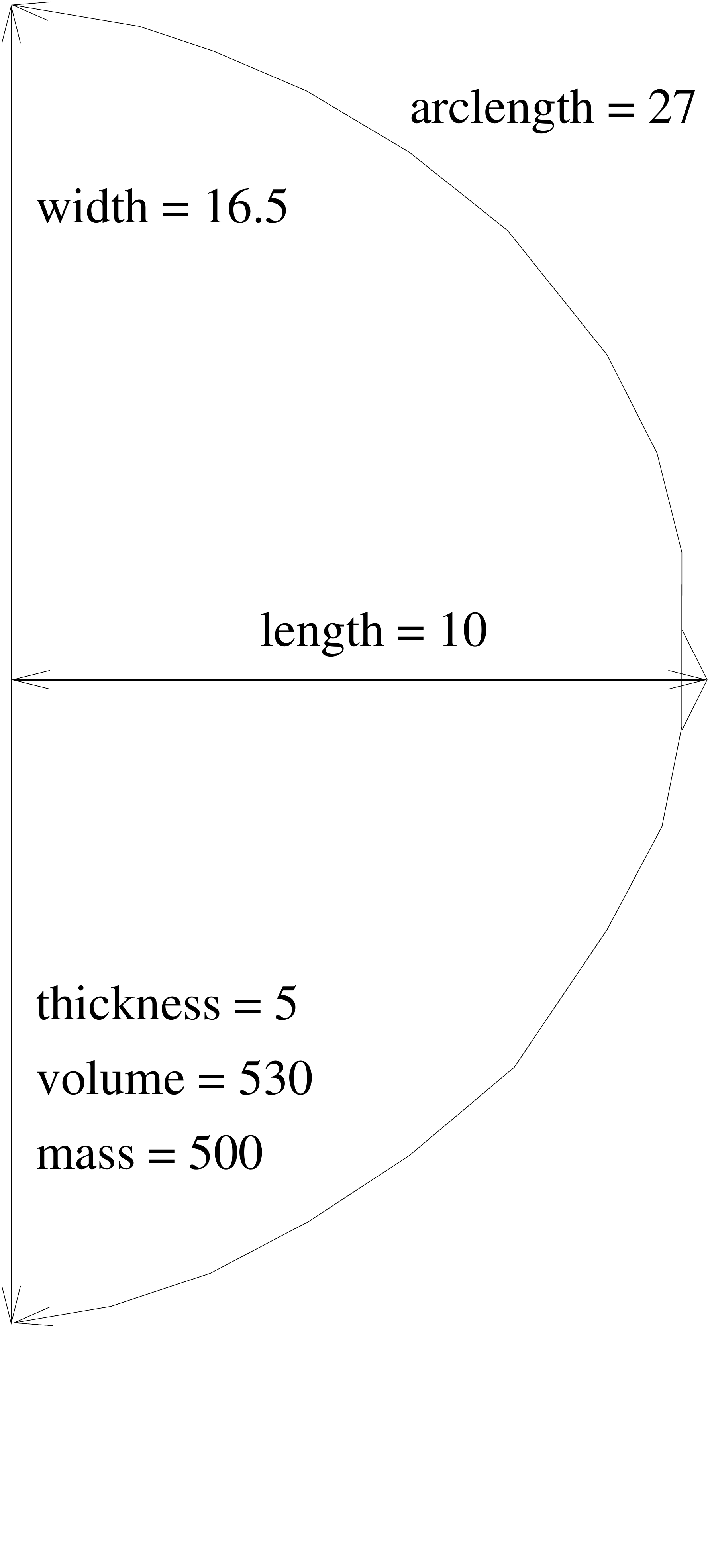}}
\centerline{Figure 10: Trajectories plotted on grid paper.\hfill Figure 11: Actual phantom dimensions.}
\ \smallskip

\begin{table}[ht!]
\caption{Coordinates Before and After CRC with $\y=Y+2.25$.}
\smallskip
\centering
\begin{tabular}{|c|r|r|r|r||r|r|r|r|}
\hline
&\multicolumn{4}{|c||}{Before}
&\multicolumn{4}{|c|}{After}\\
\cline{2-9}
Nodule & X & -Y & $\y$ & Z & X & -Y & $\y$ & Z \\
\hline
A & -1.75 &  0.75 & 1.50 & 4.75 & -2.25 &  0.00 &  2.25 & 5.75 \\
\hline
B & -0.75 & -0.50 & 2.75 & 4.50 & -1.25 & -1.00 &  3.25 & 5.50 \\
\hline
C &  1.25 &  0.88 & 1.37 & 4.75 &  1.38 &  0.25 &  2.00 & 5.50 \\
\hline
D &  1.75 & -0.50 & 2.75 & 4.50 &  2.00 & -1.00 &  3.25 & 5.63 \\
\hline
E & -2.50 &  0.00 & 2.25 & 2.25 & -3.25 & -0.75 &  3.00 & 2.63 \\
\hline
F &  0.00 &  0.13 & 2.12 & 2.50 &  0.00 & -0.75 &  3.00 & 3.00 \\
\hline
G &  3.00 &  0.25 & 2.00 & 2.25 &  3.50 & -0.30 &  2.55 & 3.00 \\
\hline
H & -3.00 &  1.00 & 1.25 & 0.25 & -3.50 &  0.00 &  2.25 & 0.50 \\
\hline
I & -1.88 & -0.75 & 3.00 & 0.00 & -2.00 & -1.63 &  3.88 & 0.38 \\
\hline
J &  0.00 &  1.13 & 1.12 & 0.00 &  0.00 &  0.50 &  1.75 & 0.00 \\
\hline
K &  0.88 & -0.75 & 3.00 & 0.00 &  1.00 & -1.75 &  4.00 & 0.38 \\
\hline
\end{tabular}
\end{table}

An easy computation shows that $E$ is residual. Its maximum entry is $0.41$ in absolute value and corresponds to the $\y$-coordinates of nodule $B$. This means 13.5\% of variation, which is a small error margin because the nodules have circa 0.5cm of diameter. 

Notice that $C$ is close to the diagonal matrix $D=1.22\,I_{3\times 3}$. In Section~\ref{lfact} we mentioned the re-scaling constant $c$ that measures the reduction in the $z$-coordinate when the breast is released from CRC to SRG, namely $h=c\,\z$. In the case of the phantom $c$ would be a little bit lesser than $1/1.22=0.82$ if it could assume the SRG-shape.

For the $Oxyz$-system one can model the compression from LAT to CRC as $(x,y,z)\to\big(kx,(-2.25-y)/k,kz\big)$ where $k=1.22$. For example, the fixed point is $(x,y,z)=(0,-2.25,0)$ on the lower plate. The point $(x,y,z)=(0,0,0)$ lies above that plate and $y$ falls from $0$ to $-1.84$ after compression.

\section{Locating Nodules for SRG}
\label{lnSRG}

Now we explain how to use our method with an example. Figure 12 was taken from the {\it Digital Database for Screening Mammography} (DDSM). For details see \url{http://marathon.csee.usf.edu/Mammography/Database.html} and \cite{DDSM}. Figure 13 shows how to mark our coordinate system on the image with colour felt-tip pens. The coordinate system was already explained in \cite{Zanka_icm2_2015}.
\\

\centerline{
\includegraphics[scale=0.4]{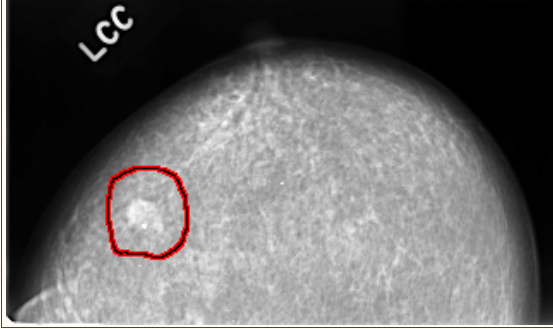}\hfill
\includegraphics[scale=0.4]{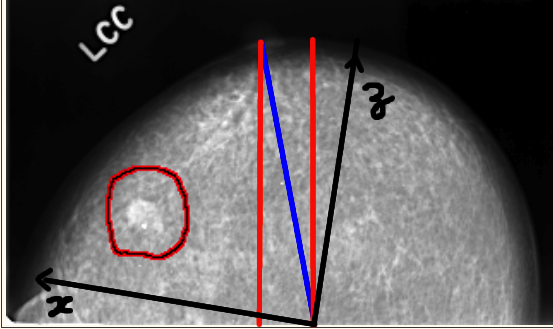}}
\centerline{Figure 12: Case 0023-1 (tumour circled in red).\hfill Figure 13: Tracing $Oxz$ with felt-tip pens.}
\ \smallskip

From Figures 12 to 13 we first draw the blue line $\b$ that is normal to the breast at the nipple. It must reach $\m$, which is the lower margin of the mammography (as shown in the picture). One can use a rule and a set square to find the correct direction of the normal. Then through the nipple we draw a red line $\r$ that is perpendicular to $\m$. Now draw $\r'$ as the parallel to $\r$ at $\b\cap\m$. With a protractor one finds $Oz$ such that $\r'$ bisects $\b\cup Oz$. Finally you can draw $Ox\perp Oz$ with the set square. 

We always use the cgs-system, as explained in \cite{Zanka_icm2_2014,Zanka_icm2_2015}. Hence units can be omitted for the sake of brevity. In that works we have used measurements of a hypotetical volunteer with a CRC very similar to the CC-view depicted in Figure 12. Therefore, we shall take the true size of Figure 12 as the one of our volunteer's CRC.

At CRC its horizontal arc measures 33, as commented in \cite{Zanka_ijser}. Since its shape is quite close to a half-circumference, then both $Ox$ and $Oz$ will cross the breast contour at $H_c:=33/\pi=10.5$. Hence in our example Figure 13 gives $(x_c,z_c)=(6.83,3.20)$, where the subscript $c$ stands for ``coordinate''. 

Now we must proceed by taking the patient's measurements described in \cite{Zanka_icm2_2014,Zanka_icm2_2015}. For the convenience of the reader we summarize them again in Figures 14 and 15.
\\

\centerline{
\includegraphics[scale=1.00]{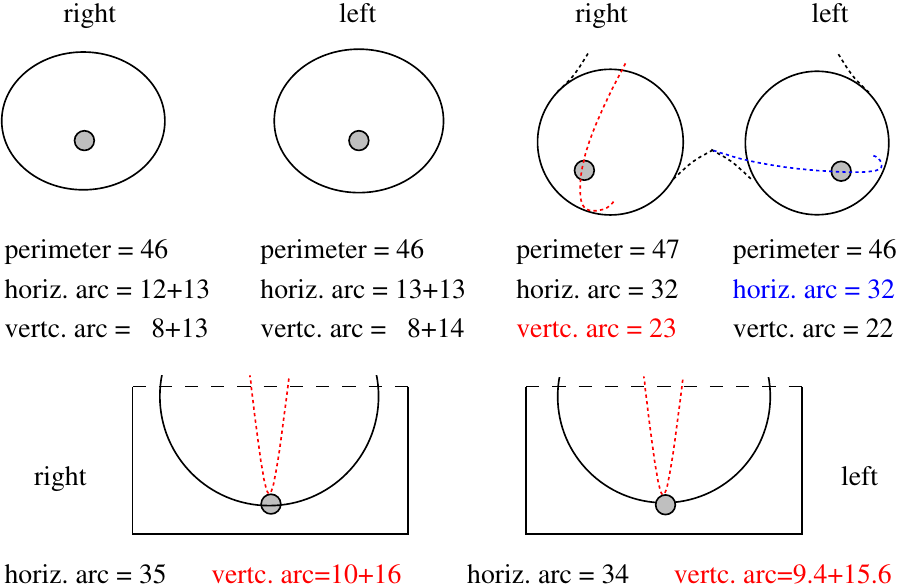}\hfill
\includegraphics[scale=1.00]{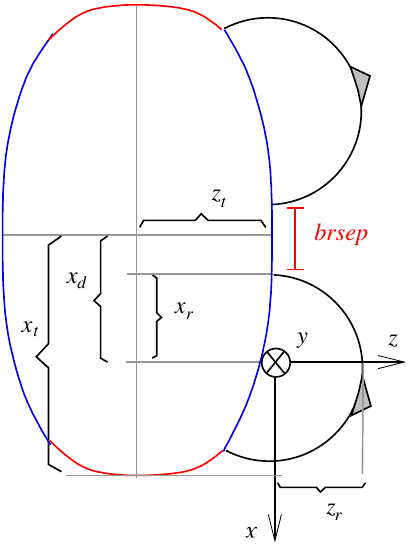}}
\smallskip
\centerline{Figure 14: Measurements of SRG, STU and LAT.\hfill Figure 15: Deduced {\it unsigned} values.}
\ \smallskip

In what follows our virtual mammographer will be applied only for the left breast. However, we recommend to measure both breasts for comparison and consistency check. As explained in \cite{Zanka_icm2_2015} we use variables like $x_r$, $y_r$ and $z_r$. In our example $y_r=z_r=22/\pi=7$ and $x_r=$ $(fthrx-brsep)/4=6.25$ at SRG. The variable $fthrx$ means ``front thorax'' and is marked as a blue arc at the bases of the breasts in Figure 15. The $z$-coordinate increases to $9.4$ at LAT and further to $H_c=10.5$ at CRC. Regarding the $x$-coordinate, it will increase to $10.82$, which is $9.17$ {\it plus} $18\%$. As explained in \cite{Zanka_ijser} we must in fact take $9.17$ at LAT, and finally $H_c=10.5$ at CRC.

According to Section \ref{exphant}, if the real breast behaves like the phantom from LAT to CRC then compression should give $(x,z)\to k(x,z)$ where $k=1.13\simeq10.5/9.3$. Moreover, from SRG to LAT we assume that both $x$ and $z$ grow proportionally with $(x_r,z_r)$. Namely, $9.4/6.25=1.5$ and $9.17/7=1.31$ to dilate $x$ and $z$, respectively. 

The nodule in Figure 13 is located at $(x_c,z_c)=(6.83,3.20)$. These values are printed with the command {\tt coors} of our simulator, together with $H_c$, and at this moment the user can change them according to another patient's. The coomand {\tt mk} deduces the nodule position at SRG as $(x_n,z_n)=(x_rx_c,z_rz_c)/H_c=(4.07,2.13)$, where the subscript $n$ stands for ``nodule''. By invoking {\tt mk} the user sees that both $L$ and $R$ are marked on the virtual breast.

Now the reader can run the first part of our simulator. It is available in the public link

\centerline{\url{https://www.copy.com/s/4fdQnwMb2IRqkPOo/ubuntu11.10.ova}}

\smallskip
\noindent
and instructions to use it are in the link {\it Softwares} of our webpage

\centerline{\url{http://www.facom.ufu.br/~nascimento}}

\smallskip
\noindent
in the PDF-file {\tt icmmps}. This file gives some general information about the first version of our simulator and includes a user manual. Follow the instructions of the manual to get SRG with the correct measurements. Afterwards mark the nodule with the command {\tt mk}. You should get Figure 16.
\\

\centerline{
\includegraphics[scale=0.39]{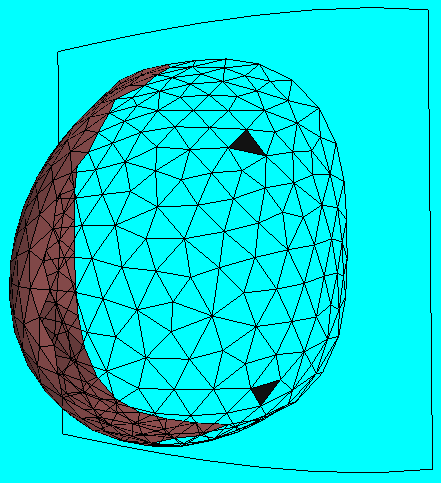}\hfill
\includegraphics[scale=0.47]{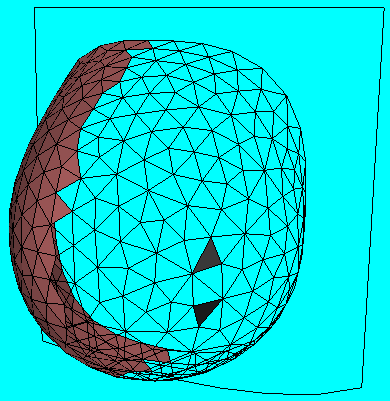}}
\centerline{Figure 16: Marking $L$ (top) and $R$ (down).\hfill Figure 17: Special case where $L\simeq R$.}
\ \smallskip

Before going ahead we just wanted to comment what our simulator will do in case $L\simeq R$ (see Figure 17). This happens exactly when $\theta$ is close to either $90^\circ$ or $270^\circ$. But then (\ref{lf}) implies that {\tt lf} is almost 100\%, namely the tumour is practically on the skin. Therefore, a simple manual examination will give its precise location.

We recall that MLO is taken after rotating the mammographer plates by $45^\circ$. The MLO view is on the plane $Ozw$, where $Ow$ is the bisectrix $y=x$ for {\it negative} values of $x$. Differently from CC, for MLO the woman is normally asked to keep her collarbones parallel to the edge of the upper plate. See the link {\it How Mammography is Performed} of

\centerline{\url{http://www.imaginis.com/mammography}}

\smallskip
\noindent
for details. 

Hence $Oz$ will be perpendicular to the lower margin of the image (see Figure 18). In fact we are interested in the points $\L$, $\P$ and $\R$, and what matters is the relative position of $\P\in\G=\arc{\L\R}$. However, you should trace $Oz$ through the highest point as depicted in our example. This point will be $(w,z)=(0,H)$.

Now the reader will promptly notice that Figure 19 does not match Figure 18 very well. This is because at MLO the patient must lift her elbow and hold a handle of the mammographer. This projects part of the breast forwards. Because of that one can see the image of the pectoralis major muscle at the lower left corner of Figure 18.
\\

\centerline{
\includegraphics[scale=0.40]{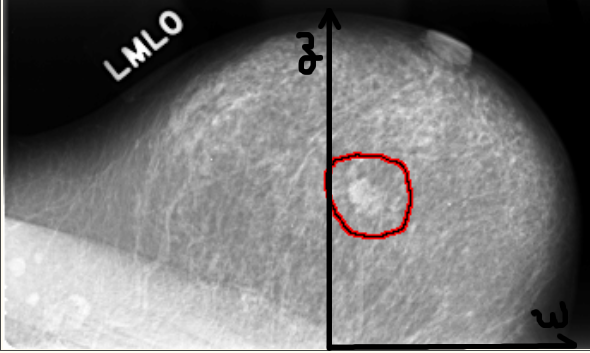}\hfill
\includegraphics[scale=0.33]{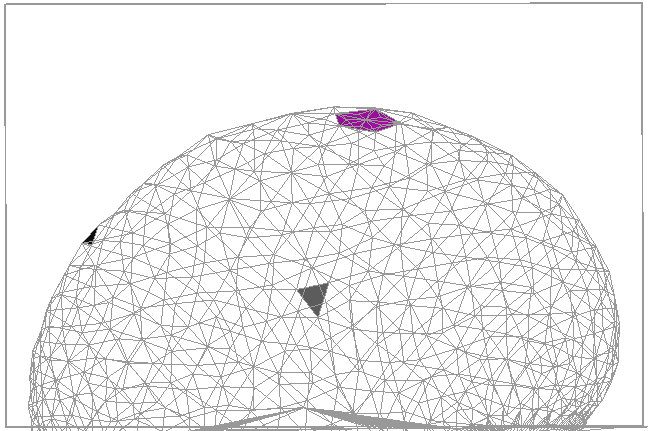}}
\centerline{Figure 18: Marking $Ozw$ on MLO.\hfill Figure 19: Comparison with the virtual MLO.}
\ \smallskip

Of course, such a movement is hard to implement computationally. Instead of doing it we resort to a well-known theorem from Complex Analysis. It characterizes all conformal automorphisms of the disk $D_H:=\{\zeta=u+iv\,|\,u^2+v^2\le H\}$. The reader does not have to learn about it. For our purposes one can simply take for granted that the function
\BE
    f(\zeta):=\frac{\ds\zeta-bH\frac{\ds b-i}{\ds 1-ib}}
              {\ds\frac{\ds1+ib}{\ds1-ib}-\frac{\ds ib\zeta}{\ds H}}
    \label{mo}
\EE
maps Figure 20 to Figure 21. The former is the upper half of $D_H$ whereas the latter is determined by the fixed point $f(H)=H$ and any chosen $b\in]0,1[$ for which $f(0)=ibH$.

We have $b=b_c/H$, where $b_c$ is the height at which $Oz$ transposes the pectoralis major muscle in Figure 18. In our example $b_c=0.95$ and $H=H_c=10.5$, thus $b=9$\%$\,=0.09$ and these are the values in (\ref{mo}) that generated Figure 21. Notice that its lower blue arc is concave because it represents the contour of the breast base, {\it not} the convex pectoralis major muscle.
\\

\centerline{
\includegraphics[scale=0.32]{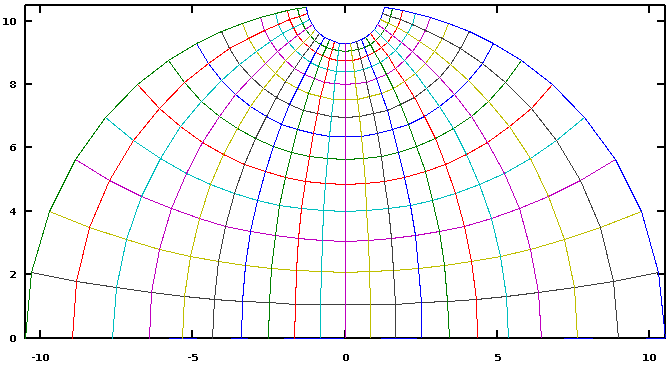}\hfill
\includegraphics[scale=0.32]{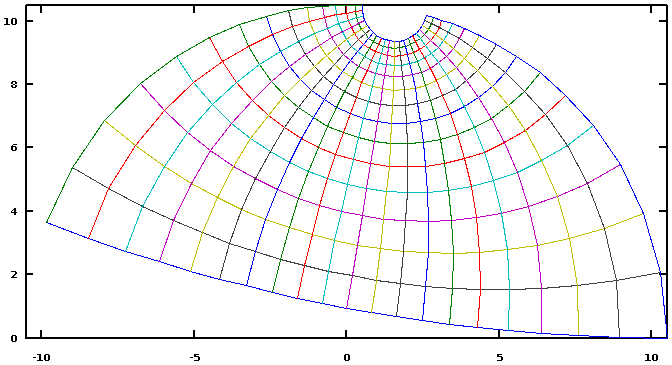}}
\centerline{Figure 20: Upper half of D.\hfill Figure 21: Its image under $f$.}
\ \smallskip

The tumour is located at $(p_w,p_z)=(1.2,4.7)$ in Figure 18. Hence its position in Figure 19 is given by $\P=f^{-1}(1.2+4.7i)=0.57+4.1i$. The command {\tt frho} prints the coordinates of the black triangles in Figure 19: $\L=-7.47+6.22i$ and $\R=-0.332+4.25i$. Technically speaking, the tumour coincides with the rightmost of these triangles. In Figure 16 it corresponds to $R$ (the lower triangle).
\\

From (\ref{rho}) this means that $P=R$, hence $\rho=1$. Now (\ref{lf}) implies that {\tt lf} is 100\%. This is just a rounding because in Section \ref{prelim} we had already mentioned that the tumour cannot be on the skin. But it is almost there, and a manual examination can easily locate it. In this case, from Figure 2 we have the coordinates $(r,p,d)$ with $d=0$. Regarding $r$ and $p$ they can be easily computed as follows.

As mentioned in Section \ref{lfact} our example from \cite{Zanka_icm2_2014,Zanka_icm2_2015} has $x_r=6.25$ and $y_r=z_r=7$. Moreover, we recall that $(x_n,z_n)=(4.07,2.13)$. Hence $h=z_n=2.13$, $a=(x_r+y_r+z_r)/3=6.75$ and
\BE
   \cos^2\theta=1-\frac{x_n^2}{a^2-z_n^2}=0.6,\label{cost}
\EE
which in our case means $\theta\simeq 127^\circ$. Notice that in Figure 9 the angle $\theta$ opens leftwards from $Oy$. If the user wants to apply the classical polar coordinates in Figure 2 ($-180^\circ\le p<180^\circ$), then $p=90^\circ-127^\circ=-37^\circ$. Finally, 
\BE
   r\simeq a\arccos\biggl(\frac{z_n}{a}\biggl)=8.44\label{appr}
\EE
but in future we shall replace the approximate formula (\ref{appr}) with a routine like the {\it virtual tape-measure} (see \cite{Zanka_icm2_2014} for details). This routine will compute the geodesic distance from the nipple numerically.

Notice that the formula in (\ref{appr}) holds {\it only} for the special case ${\tt lf}\simeq100$\%. However, the formula in (\ref{cost}) gives the correct $\cos^2\theta$ to be applied in (\ref{lf}). It is the relation $p=90^\circ-\theta$ that holds {\it only} when ${\tt lf}\simeq100$\%.

For the sake of generality let us take another $\rho$ not too close to either $0$ or $1$. Figure 19 was obtained with the command {\tt mlo} of our simulator. Now the command {\tt coors} indicate example values that the user can change according to another patient's. These are {\tt bc}, {\tt pw} and {\tt pz}. We have chosen $(p_w,p_z)=(-5.2,6.1)$ as default. The command {\tt frho} is a mnemonic for ``find rho'', and for these values it will give $\rho=0.317$ and ${\tt lf}=73$\%. This means that the tumour in Figure 18 would lie on the left hand side of $Oz$ at a point $\P\in\arc{\L\R}$ such that $\rho=0.317$ in (\ref{rho}). We have $\P=f^{-1}(-5.2+6.1i)\simeq -6.6+4.1i$. 

The reader must be curious about how we computed $|\arc{\L\P}|/|\arc{\L\R}|=\rho=0.317$, since we do not know the shapes of the arcs represented by Figure 8 left. We first consider $\Q\in\ovl{\L\R}$ such that $|\ovl{\L\Q}|/|\ovl{\L\R}|=0.317$, which is given by $\Q:=\R+(1-0.317)(\L-\R)=-5.2073+5.5955i$. By comparing Figures 22 and 23, $f^{-1}$ will carry points counterclockwise in such a way that $\Q':=f^{-1}(\Q)=-6.4193 + 3.5287i$ lies {\it below} the original segment $\ovl{\L\R}$. Notice that we do {\it not} take $f^{-1}(\L)$ and $f^{-1}(\R)$ because Figure 19 is {\it already} represented by Figure 20, whereas Figure 21 represents Figure 18.

But an arc $\arc{\L\R}$ that passes through $\Q'$ is more curved than the ones suggested by Figure~8. Hence we replace $\Q$ with $p_w+ip_z=-5.2+6.1i$ and get a less curved arc. Moreover, notice that $f$ does not deform the upper disk very much. That is why we take the equality (\ref{rho}) for granted and {\it consider} that $|\arc{\L\P}|/|\arc{\L\R}|=0.317$.

In the next section we explain how to apply the layer factor in our simulation.

\section{Applying the Layer Factor}
\label{appl}

Now we have $(x_n,z_n)$ and $\rho$. These three data will give a unique position of the tumour inside the breast as depicted in Figure 3. As a matter of fact, if we use the symmetrized breast from Figure 9 the missing coordinate $y_n$ will be
\BE
   y_n=(1-2\rho)\sqrt{a^2-x_n^2-z_n^2}.\label{wyen}
\EE
Therefore, the triple $(r,p,d)$ that will guide the surgeon to reach the nodule is given by
\BE
   r=a\arccos\biggl(\frac{z_n}{\sqrt{x_n^2+y_n^2+z_n^2}}\biggl);\label{r}
\EE
\BE
   p=\arccos\biggl(\frac{x_n}{\sqrt{x_n^2+y_n^2}}\biggl);\label{p}
\EE
\BE
   d=a-\sqrt{x_n^2+y_n^2+z_n^2}.\label{d}
\EE

In (\ref{r}) the arc-function must return values in {\it radians}, not degrees. Notice that the special cases $\rho=0$ and $\rho=1$ lead us back to (\ref{cost}) and (\ref{appr}), for $p$ and $\theta$ are then complementary. Our simulator gives $(r,p,d)$ as soon as the user invokes {\tt cnt} at the command line. In our example $(x_n,z_n)=(4.07,2.13)$ and $\rho=0.317$, hence
\BE
   (r,p,d) = \big(7.5921,23.977,1.8125\big).\label{rpd1}
\EE
But these values are still computed via (\ref{r}-\ref{d}). Because of that, the virtual mammographer will print (\ref{rpd1}) as an ``a priori'' location. In future we shall implement routines that will give $(r,p,d)$ according to the asymmetric shape.

Anyway, we recommend the user to proceed with the simulation. It will now include the layer and the virtual nodule, as depicted in Figures 3 and 4 of Section~\ref{prelim}. Eventually they will end up with the virtual CRC depicted in Figure 22. But for this one the black triangle has coordinates
\BE
   (x_c,z_c)=(4.94,2.82),\label{lf1}
\EE
which are considerably different from the original $(6.83,3.20)$. Since we have been using approximations, the user can try another layer factor, for istance 81\%. This will give Figure 23.
\\

\centerline{
\includegraphics[scale=0.257]{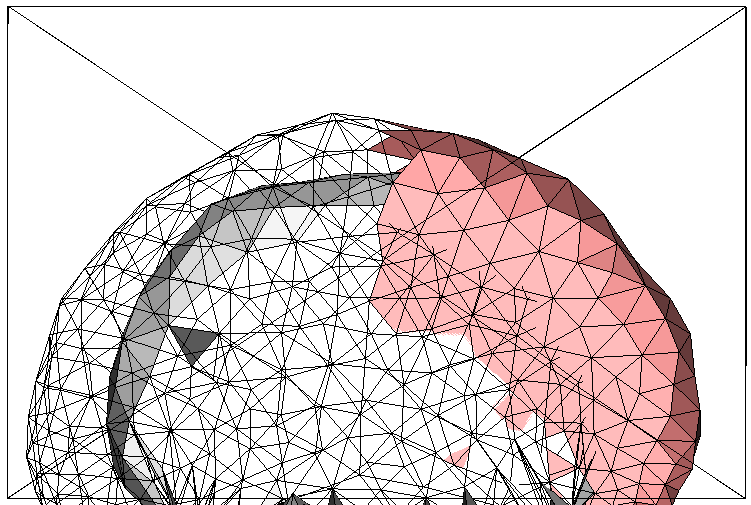}\hfill
\includegraphics[scale=0.270]{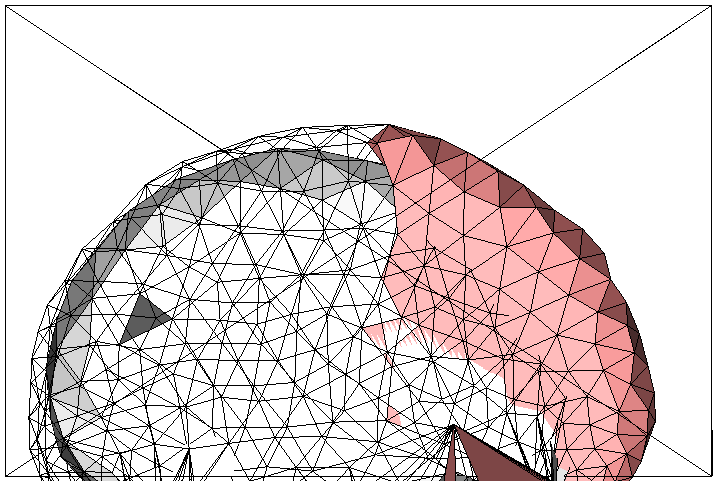}}

\centerline{Figure 22: Checking {\tt lf} at 73\%.\hfill Figure 23: Checking {\tt lf} at 81\%.}
\ \smallskip

For ${\tt lf}=0.81$ our simulator will print
\[
   (x_c,z_c)=(5.95,3.26),
\]
which is pretty closer to the original $(6.83,3.20)$ than (\ref{lf1}). Moreover, the new $\rho=0.2$ now gives
\BE
   (r,p,d) = \big(7.9024,36.096,1.2812\big).\label{rpd2}
\EE

Of course, now we must simulate MLO. The virtual mammographer will give a picture very similar to Figures 22 and 23. Morover, it will print 
\BE
   (p_w,p_z)=(-5.83,4.79),\label{lf2}
\EE
which is distant from the original $(p_w,p_z)=(-5.2,6.1)$ by less than 1.5cm, as in the previous view. However, we recall that our original $(p_w,p_z)$ from Section~\ref{lnSRG} is hypotetical because Figure 18 shows a nodule that technically gives $\rho=1$. Since this is too extreme for a general example then we {\it guessed} $(-5.2,6.1)$.

In order to improve accuracy one should start a convergence process: replace the inital guess with another point closer to (\ref{lf2}); simulate again to find new values of {\tt rho} and {\tt lf}; apply these new values in the simulator; get a new picture like Figure 22; change {\tt lf} to approach the original $(x_c,z_c)=(6.83,3.20)$ at CRC, as depicted in Figure 23; simulate MLO and get new values at (\ref{lf2}). Repeat this whole processes until obtaining the desired precision.

However, we consider that an error margin of 1.5cm is acceptable. The surgeon will then try (\ref{rpd2}) as illustrated by Figure 6 of Section~\ref{prelim}. Once the point $(r,p)\simeq(7.9,36^\circ)$ is marked on the breast the scalpel will cut till depth $d\simeq 1.3$cm.

At the end of Section~\ref{lfact} we mentioned that the surgeon can check consistency between $p$ and $\theta$ by deducing this latter. This makes sense if {\tt lf} is close to 100\%. Hence, in our example we expect that the deduced $\theta$ will be around $50^\circ$.

\section{Conclusions}
\label{concl}

As explained at the Introduction, along the decades there has been an effort to improve Breast-Conserving Surgery (BCS). For this purpose one of the crucial tasks is to minimize the size of breast portion for removal. Therefore, one seeks the exact position of the tumour inside the breast. 

The results presented herein are a valuable help for BCS. In order to test them one should start with real patients for whom a mastectomy was prescribed. Then Sections \ref{prelim}, \ref{lnSRG} and \ref{appl} will guide the medical staff as part of the surgical preparation. Finally, if the tumour is really found this will validate our method, in spite of the simplifications that we have been using. 

Some of them were already mentioned in the previous sections, but here we list the most relevant imprecisions of our method:

\begin{description}[leftmargin=2.5em,style=nextline]
\itemsep0pt\parskip0pt\parsep0pt
\item[ \ {\bf 1.}] We have used Sections \ref{lfact} and \ref{exphant} to {\it assume} that the coordinates $(x,z)$ deform linearly from SRG to CRC, namely $(x,z)\to(1.68x,1.5z)$ in our example.
\item[ \ {\bf 2.}] Figure 8 does not consider the thickness of the compressed breast at CRC and MLO, as depicted in Figure 4.
\item[ \ {\bf 3.}] In practice, the layer-approach was not proved to be a good approximation yet.
\item[ \ {\bf 4.}] Some of our formulas still rely on the symmetrized breast at SRG.
\item[ \ {\bf 5.}] We do not know if the patient 0023-1 was correctly positioned when the X-ray images were taken.
\item[ \ {\bf 6.}] Once a mastectomy is prescribed, the measurements of the patient's breast must consider variations in the body that might have happened after the mammographies were taken. For example: loss or gain of weight, breast swelling due to menstrual cycle, etc.
\end{description}

We can take {\bf 1} to {\bf 4} into account for future improvements in our simulator. However, {\bf 5} and {\bf 6} will always depend on the user's carefulness at applying our method correctly.

\section*{Acknowledgements}

The second author thanks his wife Clarice for the motivation that she gave to develop this research. We are grateful to Prof Jos\'e Artur Quilici Gonzalez, Federal University of ABC, for his assistance with the state of the art. We thank Dr Ana Cl\'audia Veronesi for details about gynaecological surgery, and her husband Alejandro Montepeloso for translating some technical terms. We also thank {\it Ribeir\~ao Pires Hospital (SP - Brazil)} for their assistance and for the concession to use their mammographer in our compression experiments with the phantom. This research is supported by FAPESP proc.No.12/16244-3.

\bibliographystyle{plain}
\bibliography{amm}
\end{document}